\documentclass[%
 reprint,
superscriptaddress,
 amsmath,amssymb,
 aps,
]{revtex4-1}

\usepackage{graphicx}
\usepackage{dcolumn}
\usepackage{bm}

\usepackage{float}

\def\n{\boldsymbol{n}}
\def\r{\boldsymbol{r}}
\def\half{\textstyle \frac{1}{2}}

\begin{document}


\title{Curvature by design and on demand in liquid crystal elastomers}%

\author{B. A. Kowalski}
\affiliation{%
Air Force Research Laboratory,  Wright-Patterson Air Force Base, OH, USA}%
\affiliation{%
Azimuth Corp., Beavercreek, OH, USA}%
\author{C. Mostajeran}
\affiliation{%
Department of Engineering, University of Cambridge CB2 1PZ, UK}%
\author{N. P. Godman}
\affiliation{%
Air Force Research Laboratory, Wright-Patterson Air Force Base, OH, USA}%
\author{M. Warner}
\affiliation{%
Cavendish Laboratory, University of Cambridge, Cambridge CB3 0HE, UK}%
\author{T. J. White}
\affiliation{%
Air Force Research Laboratory,  Wright-Patterson Air Force Base, OH, USA}%

\date{\today}

\begin{abstract}
The shape of liquid crystalline elastomers (LCEs) with spatial variation in the director orientation can be transformed by exposure to a stimulus.  Here, informed by previously reported analytical treatments, we prepare complex spiral patterns imprinted into LCEs and quantify the resulting shape transformation.  Quantification of the stimuli-induced shapes reveals good agreement between predicted and experimentally observed curvatures.  We conclude this communication by reporting a design strategy to allow LCE films to be anchored at their external boundaries onto rigid substrates without incurring internal, mechanical-mismatch stresses upon actuation, a critical advance to the realization of shape transformation of LCEs in practical device applications.
\end{abstract}

\maketitle


Liquid crystal elastomers (LCEs) are exciting materials that show promise for enabling multifunctional character in flexible devices.  The mechanical response of these materials is inherently and programmably anisotropic, governed by the molecular-level liquid crystalline orientation.  This generates a rich variety of reversible, shape-morphing behaviors \cite{kowalski2017pixelated} with large strains (of order 400\%) and actuation force that can rival human muscle \cite{ ware2015voxelated}.
Recent work has advanced the capabilities for rapid, high-resolution spatial patterning of the director profile in LCE films \cite{ kowalski2017voxel}.  In general, the intended director pattern is inscribed onto an intermediary alignment layer using any of several of non-contact techniques including linear photopolymerization \cite{yaroshchuk2012photoalignment}, photo-reorientation of azo dye molecules \cite{ ware2015voxelated,guo2016high}, or lithographic inscription of micro-grooves \cite{xia2016guided}.   The alignment layer then enforces self-assembled orientation of the liquid crystal monomer and oligomeric precursors which are drawn into the alignment cells and subsequently polymerized to form a free-standing monolithic film.

Exposing LCEs to various stimuli (e.g. heat, light, or solvent) then generates a locally programmed, anisotropic mechanical response, resulting in the conversion of 2D flat films to complex 3D shapes \cite{ahn2016photoinduced}.  The diversity of the accessible shapes continues to open up new potential applications including haptic devices \cite{torras2014tactile}, flow control surfaces, microfluidics, and deployable antennas, among others (for an overview, see \cite{leng2011shape}).
A considerable fraction of the existing literature, both analytical and experimental, has explored elemental building blocks, such as cantilevers, cones, and in some instances arrays of these component units to form more complex shapes \cite{modes2011blueprinting,aharoni2014geometry}. Some recent efforts have pursued the realization of smoothly varying curved structures \cite{aharoni2014geometry,mostajeran2015curvature,mostajeran2016encoding,plucinsky2016programming}.  Further,  one subject of practical significance is the inverse design of a director profile that will generate a desired curvature.  A general solution to the inverse design problem is proposed in \cite{aharoni2014geometry}, which uses double-sided patterning of elastomeric films to inscribe the second fundamental form of the target surface. If one restricts to single-sided patterning, whereby an in-plane director field is prescribed without variations along the thickness of the sample, the inverse problem becomes more challenging as a degree of freedom is removed from the design parameters. Indeed, for most target surfaces a solution does not exist to this form of the inverse problem.
Although specific examples have been calculated in \cite{aharoni2014geometry,mostajeran2015curvature,mostajeran2016encoding}, in the absence of generally available, concrete solutions to realize shapes of interest, there has been a reliance on intuitive design or on iterative approaches such as topology optimization \cite{fuchi2015topology,dunn2009photomechanics,gimenez2017modeling}.

Here, we focus on a subset of director profiles for which analytical solutions exist.  Multiple families of such profiles have been proposed by theory, with only a few thus far subjected to experimental examination \cite{mostajeran2016encoding}.   In this way, this communication is a closed-loop exploration derived from analytically predicted patterns, realized in LCE films enabled by advances in materials and processing methods, and characterized by optical scanning.  We show herein that the desired director profiles are indeed realizable in LCE films, and that the resulting curvature quantitatively matches the shape predicted \textit{ab initio}.  The utility of this theory-led design approach is extended to the fabrication of anchorable actuating films whose perimeters can be affixed to rigid substrates.

\begin{figure}
\centering
\includegraphics[width=1\linewidth]{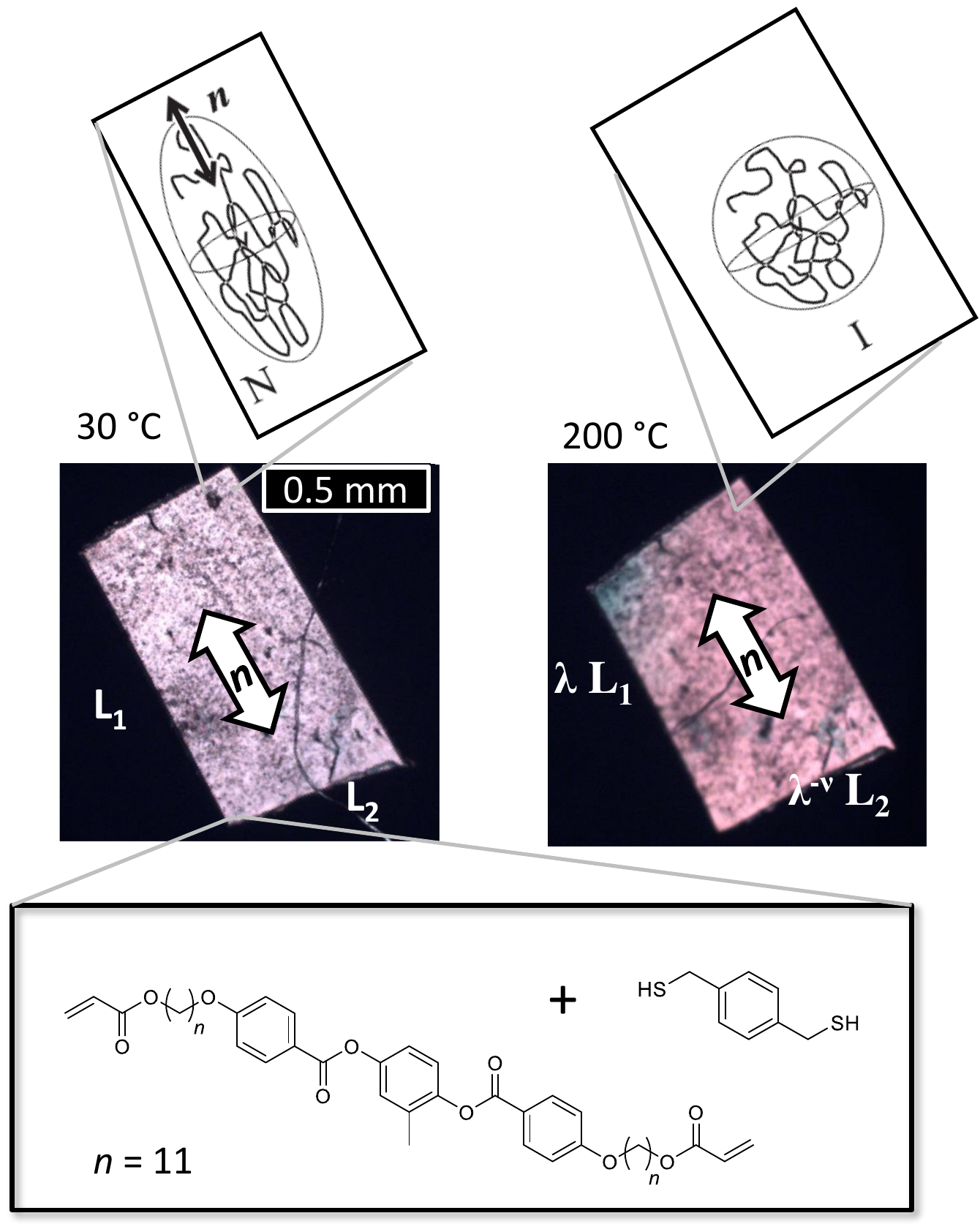}
  \caption{(a) Polarized optical micrographs of an LCE floating in a bath of silicone oil.  Upon heating, the microscopic order parameter is reduced (inset) leading to a macroscopic deformation along the director. (b) Chemical structures of monomers used here, following \cite{godman2017liquid}.}
\label{fig:1}
\end{figure}

We consider thin LCE films with nematic planar alignment (i.e. the director is everywhere in the plane of the film, with only the azimuthal angle varying spatially).  In response to external stimuli (heat, light, etc.), the order parameter is decreased, and the average polymer chain configuration converts from a prolate spheroid to spherical \cite{warner2003liquid} (Figure \ref{fig:1}). On the macroscopic scale, this translates into a contraction along the director by some factor $\lambda$, partially compensated by an expansion in other directions, by a factor $\lambda^{-\nu}$ where $\nu$ is the thermo-optical analogue to the Poisson ratio \cite{modes2011blueprinting}.

The deformed shape is predicted by exploring non-Euclidean metrics as arising in films with residual stress \cite{efrati2013metric}.  Following \cite{warner2003liquid},  the spontaneous local deformation tensor takes the form
\begin{equation} \label{deformation}
F=(\lambda-\lambda^{-\nu})\n\otimes\n+\lambda^{-\nu}I
\end{equation}
where $I$ is the identity operator and $\n$ is the director. Here, we make the further assumption that $\lambda$ and $\nu$ are constant (e.g. no differential swelling), and that the only spatial variation is in the director orientation, that is $\n(\r)$. The two parameters $\lambda$ and $\nu$, along with the prescribed director pattern $\alpha(\r)$, are the only inputs to the model. Both $\lambda$ and $\nu$ can be estimated independently, here by examining the thermal response of an LCE film with uniform planar alignment in a temperature-controlled oil bath.  As the bath is slowly heated, the dimensional changes in the sample are visually apparent (Figure \ref{fig:1}).

Here, we use a recently reported approach \cite{godman2017liquid} to prepare LCEs that exhibit large deformations upon relatively modest temperature changes.  The synthesis of these materials is largely based on commercially available liquid crystalline monomers (diacrylates) that homopolymerize to form glassy liquid crystalline films with good uniformity \cite{broer1989situ}. To these materials, a minimal amount of chain transfer agent has been shown to dramatically reduce the crosslink density and yield LCEs with actuation strain of nearly 200\% and soft elasticity that rivals polysiloxane LCE (Figure \ref{fig:1}). This chemistry is compatible with high-resolution photoalignment techniques as described in \cite{ ware2015voxelated,kowalski2017voxel}.

\begin{figure*}
\centering  
\includegraphics[width=1\linewidth]{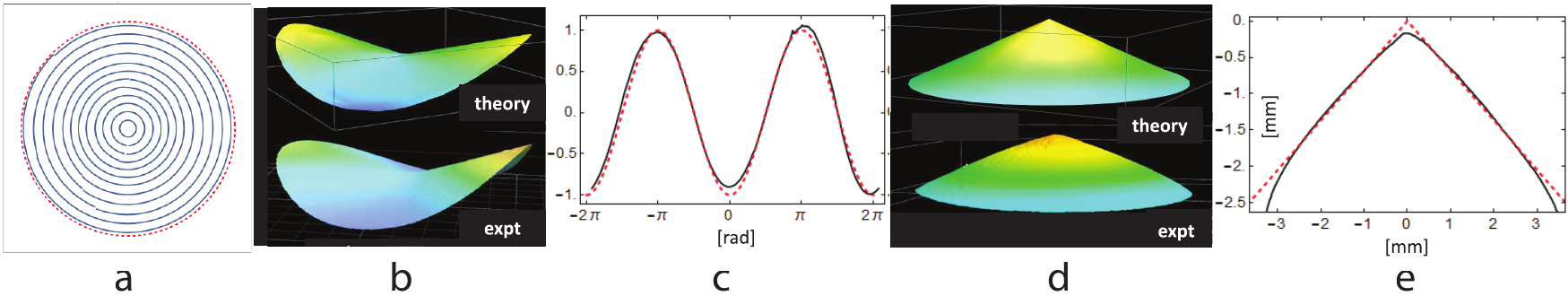}
  \caption{(a) Azimuthal (+1 defect) director profile templated into an LCE film. This profile produces deformation into either (b) negative Gaussian curvature upon cooling to 30C or (d) positive Gaussian curvature upon heating to 150C; at the reference temperature $T_0 \sim 95C$, the film is flat. Good quantitative agreement is observed between predicted and measured shapes.  This is confirmed by height profiles (c,e), with measured heights shown by solid black traces and \textit{ab initio} predictions shown by dashed red traces.  Profiles are measured around the perimeter of the anti-cone (b,c) and in a slice through the center of the cone (d,e).  For simplicity, no angular averaging is performed.}
  \label{fig:2}
\end{figure*}

We now consider a class of radially-symmetric director profiles containing a point defect of topological charge $m=+1$, appropriate strictly only to 2D director fields, in which the director makes an angle $\alpha(r)$ with the radial direction.
The two limiting cases are pure azimuthal orientation  ($\alpha=\pi/2$) and pure radial orientation ($\alpha=0$). Preparing a glassy liquid crystalline polymer network (LCN) or LCE with the azimuthal pattern (Figure \ref{fig:2} a) generates a conical deformation upon exposure to thermal or photonic stimuli \cite{mcconney2013topography}. By a simple geometric argument, the cone half-angle is predicted \cite{modes2010first_cones} to be:
\begin{equation}
\phi=\arcsin \left(\lambda^{1+\nu}\right).
\end{equation}
As shown in Figure \ref{fig:2}, the fabricated LCE film, when heated above its reference temperature $T_0$, forms a cone that closely matches this theoretical prediction, in magnitude as well as shape.  In this and the following experiments, the fabricated circular film of diameter 1 cm and thickness 30 microns is free-standing upon a temperature-controlled glass substrate with non-stick coating.  \textit{In situ} characterization of the deformed shapes is performed with a structured-illumination optical scanner (Keyence VR-3200) which provides a quantitative 3D height map with micrometer height resolution. In Figure \ref{fig:2}c, the film is heated from $T_0 \sim 95C$ to 150C, corresponding to $\lambda \sim 0.88$ in the independent oil-bath measurement discussed above.  The value of $\nu$ is assumed to be exactly 0.5, corresponding to perfect volume conservation, and consistent with the above measurement.

The most severe discrepancy in shape is observed at the cone tip.  The predicted ideal cone shape has infinite bend at the tip, so that the elastic bend energy cost exceeds the otherwise dominant stretch cost. To accommodate this, the sheet instead forms a rounded cap, over a region whose size is of the same order as the film thickness, as discussed in \cite{modes2011gaussian}.

Here we can also confirm another prediction made in \cite{modes2011gaussian} that cooling the LCE film below $T_0$ produces a curvature of the opposite sign, referred to as an anti-cone (in accordance with the orthogonal duality result of \cite{mostajeran2016encoding}).  We also note good agreement with the closed-form prediction of the anti-cone shape laid out in \cite{modes2011gaussian}:
\begin{equation}
h(r,\phi)=A r \sin(n\phi)
\end{equation}
with $n=2$ as the lowest-energy mode.

\begin{figure}
\centering  
\includegraphics[width=1\linewidth]{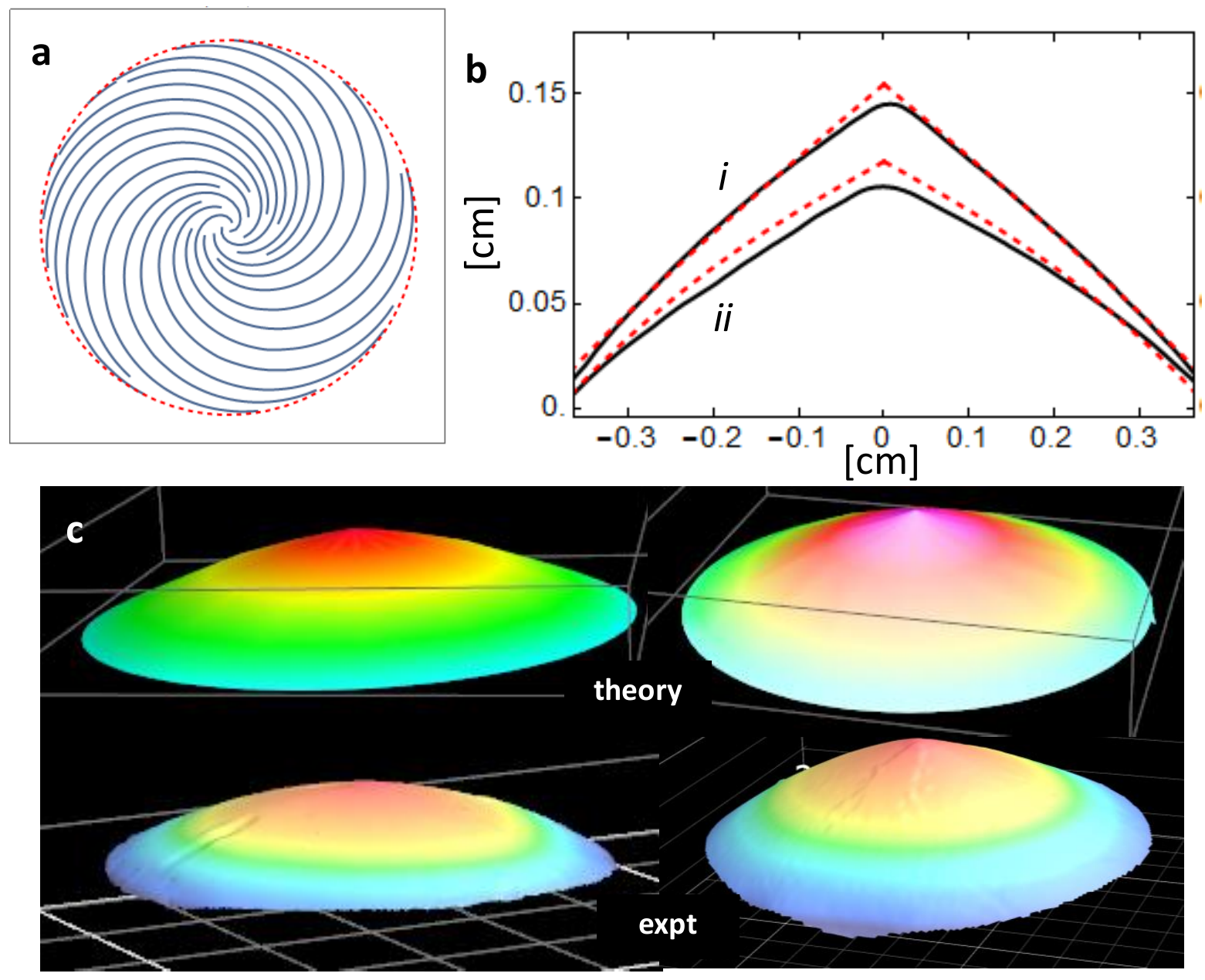}
  \caption{Spiral director profiles predicted to yield constant Gaussian curvature $K > 0$ upon exposure to a thermal stimulus.  (a) Representative director profile, with the spiral described by parameter $c = - 0.35$.  (b) Predicted and measured height profiles for two values of $c$: (i) $c = - 0.75$ and (ii) $c = - 0.35$, showing fine control of the realized curvature.  (c) Predicted and measured 3D shapes for the same two profiles $c = - 0.75$ (right) and $c = - 0.35$ (left).}
  \label{fig:3}
\end{figure}

A broader and more tailorable range of radially symmetric shells arises by extending \cite{mostajeran2016encoding} to a radial variation in  $\alpha(r)$. Calculating the metric components with respect to polar coordinates, we arrive at the following equation for the Gaussian curvature $K=K(r)$ in terms of $\alpha=\alpha(r)$:
\begin{equation} \label{GC alpha}
K
=\frac{\lambda^{-2}-\lambda^{2\nu}}{2}\left[\left(\alpha''+\frac{3}{r}\alpha'\right)\sin
(2\alpha)+2\alpha'^2\cos (2\alpha)\right]
\end{equation}
where $'$ denotes differentiation with respect to $r$. Additionally considering in-material length changes, both circumferentially and in surface trajectories that become radii on deformation, one can determine the curvatures of the evolving surfaces \cite{mostajeran2016encoding}.
Of particular interest is the case of surfaces of revolution of constant and positive Gaussian curvature, that is, spherical spindles where the
two principal curvatures are typically not equal and are not constant \cite{mostajeran2016encoding}. In the case where the principal curvatures agree, the spherical spindle reduces to a spherical cap.
On solving eqn~(\ref{GC alpha}) as a differential equation for $\alpha(r)$, constancy of $K$ is satisfied by profiles with the form:
\begin{equation}  \label{profile1}
\alpha(r)=\half\arccos\left[ -\left(\frac{1+c}{L^2}\right)r^2+c\right]
\end{equation}
where $0 \leq r \leq L$ and $c$ is a tuning parameter that determines curvatures and relates to the in-material lengths mentioned above. $L$ is a scaling factor that is determined by the dimensions of the elastomeric film samples used in the patterning process.
The expression in eqn~(\ref{profile1}) is a reformulation of director fields from \cite{mostajeran2016encoding} to ensure optimal scaling for the utilized patterning process.
Various magnitudes of constant positive curvatures $K$ can be realized through appropriate choice of the parameter $c$ according to
\begin{equation}
K=\frac{2(1+c)(\lambda^{-2}-\lambda^{2\nu})}{L^2}.
\end{equation}
As presented in Figure \ref{fig:3}, even for more complicated director profiles, this \textit{ab initio} prediction of $K$ is still in good quantitative agreement with measurements.  In other words, for given material parameters $\lambda$ and $\nu$, the desired apex angle of the actuating spindle can be precisely tailored through choice of the parameter $c$. The theoretically predicted surfaces of Figure \ref{fig:3} (c) used in the quantitative analysis of the experimental results were computed using the parametrization given in equations (3.19) and (3.20) of  \cite{mostajeran2016encoding} with the coefficients determined by the experimental parameters utilized here.

  For the two samples shown here, heating is from $T_0 \sim 95C$ to 120C, corresponding to a measured $\lambda = 0.95$, with $\nu$ again taken to be 0.5 exactly.  As before, the greatest discrepancy in shape occurs where the predicted sharp central tip is replaced by a smooth rounded dome, over a length comparable to the 30 micron film thickness.

\begin{figure}
\centering
\includegraphics[width=1\linewidth]{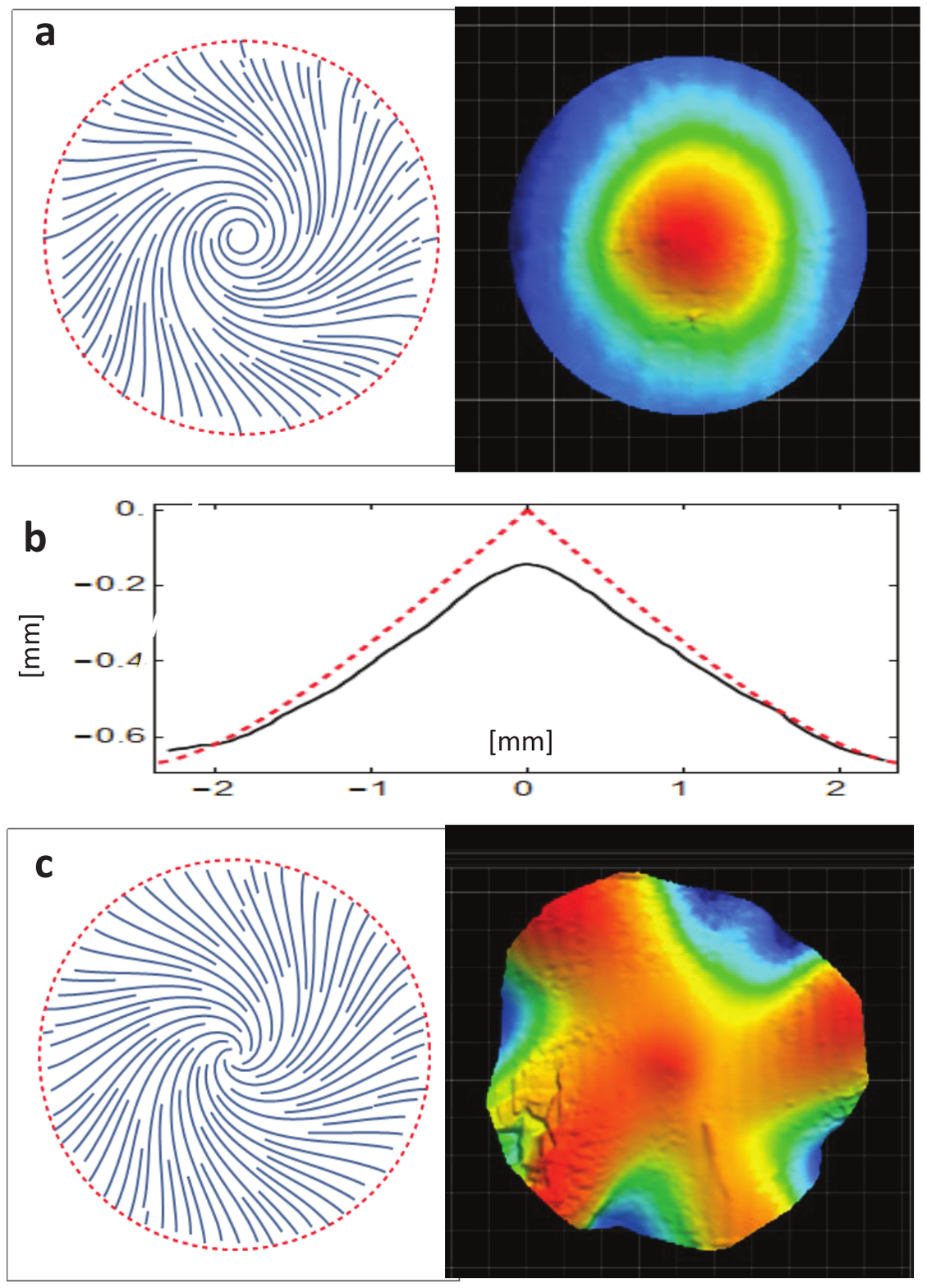}
  \caption{The negative-curvature equivalent of the previous family of director profiles.  (a) Profile with $c = - 1$ and resulting measured height profile showing the expected hyperbolic spindle.   (b) Representative radial slice through (a), showing good agreement between the predicted (dashed line) and measured (solid line) profile.  (c) Profile with $c =  0.35$, which exhibits symmetry breaking ruffles as expected from theory for this choice of $c$. All other parameters are as in Fig. 3.}
    \label{fig:4}
\end{figure}

The preceding analysis can be repeated for the condition of constant but negative Gaussian curvature.  This condition is satisfied by:
\begin{equation}
\alpha(r)=\half\arccos\left[\left(\frac{1-c}{L^2}\right)r^2+c\right].
\end{equation}
Again, the choice of parameter $c$ determines the curvature $K$, now as follows:
\begin{equation}
K=-\frac{2(1-c)(\lambda^{-2}-\lambda^{2\nu})}{L^2}.
\end{equation}
This will have a closed-form, radially symmetric solution in the form of a hyperbolic spindle (i.e. surface of revolution of constant negative Gaussian curvature) provided that
\begin{equation}  \label{ruffle c}
c < 1 - \frac{2}{1+\lambda^{1+\nu}},
\end{equation}
as shown in  \cite{mostajeran2016encoding}. For choices of $c$ outside of this range, symmetry-breaking ruffles will inevitably form due to an excess in circumferential length at any geodesic radial distance from the central topological defect, just as in the anti-cone case discussed above.
As presented in Figure \ref{fig:4},  agreement with this theoretical prediction is observed in experiments, where the choice of $c = -1$ satisfying (\ref{ruffle c}) is shown to produce a surface of revolution of negative Gaussian curvature, whereas the choice $c = 0.35$ which does not satisfy (\ref{ruffle c}) results in symmetry breaking. Moreover, in the case of $c = -1$, the realized shape of the hyperbolic spindle exhibits good agreement with \textit{ab initio} predictions except near the center, where bend energies become non-negligible due to the sharpness of the tip.

Finally, we conclude by considering the important practical challenge of realizing ï¿½anchorableï¿½ films.  That is, materials prepared from patterns for which the circumference of the film's outer rim can have a null deformation at particular $\lambda$s as the whole is deforming.  Accordingly, LCE films can be anchored to rigid substrates without incurring stress or delamination at the perimeter for these $\lambda$s.   Anchoring shape-morphing LCE films is expected to be critical in allowing the integration of these materials into a range of proposed applications such as haptic displays \cite{torras2014tactile},  microfluidic pumps  \cite{chen2011photodeformable},  or reconfigurable optical devices \cite{schuhladen2014iris}.  A few anchorable profiles have been proposed in [14].  However, these patterns rely on the exotic property $\nu > 1$ (sometimes found in nematic glasses), whereas LCEs characteristically exhibit volume-conserving deformations corresponding to $\nu \sim  0.5$.

We begin by noting that the director is everywhere oriented in the plane of film, no matter how its azimuthal angle varies with position; this constraint is imposed by the nature of the alignment techniques used here. The overall in-plane surface area of a film upon heating must change by a factor: $\lambda \times \lambda^{-\nu}= \lambda^{1-\nu}< 1$ since $\lambda < 1$.  In the usual volume-conserving case, this represents a reduction in overall area, along with a corresponding gain in out-of-plane thickness.  But if the perimeter of the film is to be held fixed, then any deformation away from the flat (drum-head) configuration requires an \emph{increase} in overall area.

However, this incompatibility can be resolved if the film reaches its deformed state by cooling rather than heating.  That is, the deformed (higher surface area) configuration corresponds to the lower temperature, and the flat (small surface area) configuration corresponds to the higher temperature.  Furthermore, the previously discussed analytical results for achieving particular shapes can still be applied here, simply by invoking a key result of  \cite{modes2011blueprinting} that the roles of heating and cooling can be reversed by replacing the director with its orthogonal dual.

\begin{figure}
\centering
\includegraphics[width=1\linewidth]{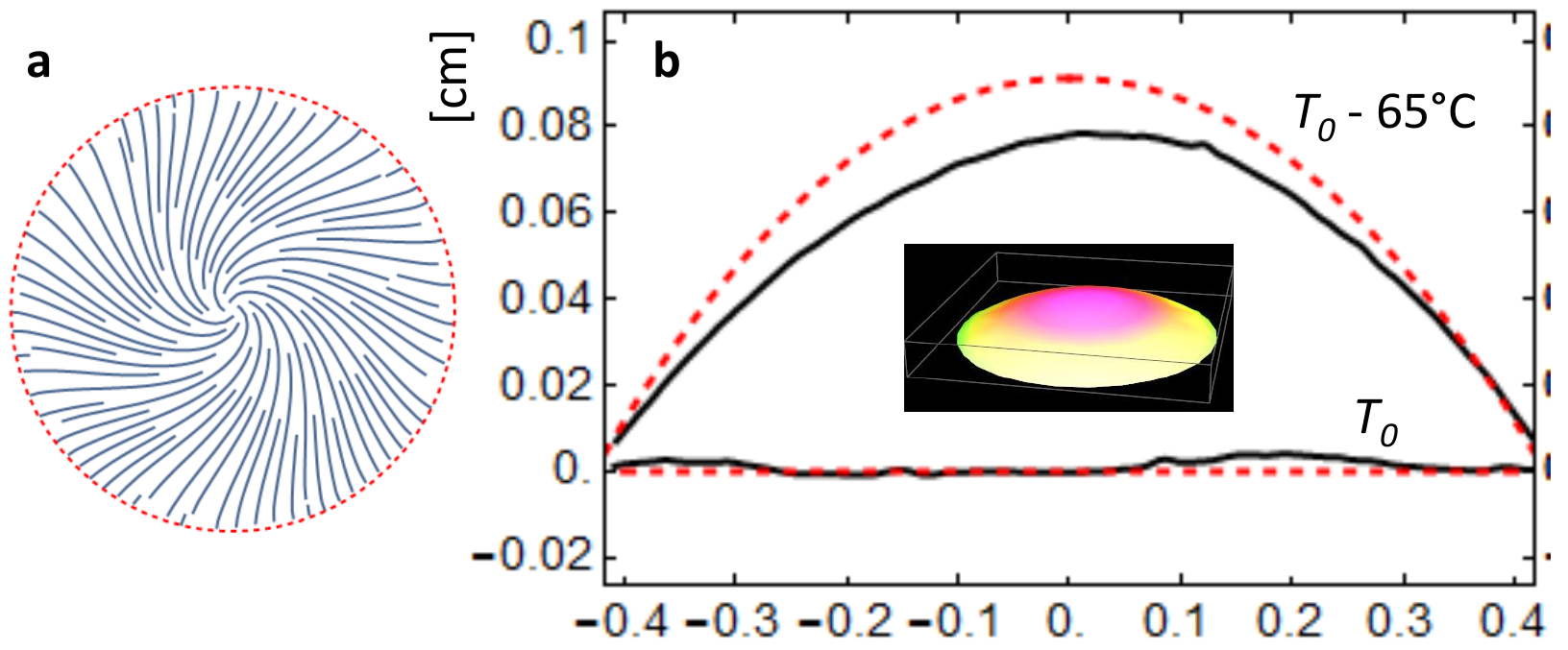}
  \caption{Anchorable spherical cap that forms upon cooling.  (a) Prescribed director profile. (b) Fabricated film, with rim glued to a metal washer to enforce fixed boundary condition. (c) Predicted (dashed line) and measured (solid line) shape, first at the flatness temperature $T_0$ and then upon cooling to room temperature.  The heating case is not shown, since the induced negative curvature is incompatible with the fixed boundary condition.}
    \label{fig:5}
\end{figure}

As a specific demonstration, we choose to fabricate an anchorable spherical cap, since this curvature may be of general interest, for example in optical devices.  The positive-curvature spiral form of the director profile discussed above is predicted \cite{mostajeran2016encoding} to yield an exactly spherical cap for the following value of the parameter $c$:
 \begin{equation}
 c=1-\frac{2}{1+\lambda^{1+\nu}}.
 \end{equation}
 The anchoring radius, $r_0$ for spherical caps is given in eqn~(3.29) of the same paper. We use the orthogonal dual of this pattern, and cool from the flat temperature ($T_0$) instead of heating.  To demonstrate anchorability to a rigid substrate, the perimeter of the film is glued to a metal washer.  Upon cooling from 95C to 30C, corresponding to an effective $\lambda \sim 0.98$, the desired deformation is achieved, without tearing or wrinkling evident on the LCE material anywhere in the film. See Figure \ref{fig:5}.

These examinations also afford insight into the more general question of whether LCEs, with their low crosslink density and extremely large deformations, can exhibit the same precise and repeatable shape changes as more highly crosslinked glassy liquid crystal networks.  In particular, LCEs exhibit ï¿½soft elasticï¿½ responses to stress within which the director can reorient.  It has remained an outstanding question whether soft elasticity would prevent realization of expected shape transformation \cite{warner2003liquid}.  In \cite{mostajeran2016encoding}, it was argued that such distortions should be negligible since the actuated film adopts a shape that has zero stretch with respect to the new metric.  This argument is corroborated by the detailed characterization of shape reported here for the first time.

In summary, a range of precisely tailored curvatures were realized in LCE films, exhibiting good quantitative agreement with analytical predictions.  This result illustrates the promise of a design approach that uses smoothly varying director profiles inspired by analytical calculations.  The cases demonstrated here represent only a sampling of a much larger design space.  We hope this initial examination of determinate problems will spur other further efforts to develop analytic solutions and advancements in the realization of inverse design models.

Finally, we show that an additional degree of design freedom is available: deformation can occur either on heating or on cooling, with the reference (flat) temperature readily tuned by adjusting the composition as reported in \cite{godman2017liquid}.  This freedom can be exploited to create deforming membranes that keep the same shape at their rim so that they can be anchored to rigid substrates with no stress penalty.  We believe this is a critical advance and hope this result enables the incorporation of LCEs into device applications.



\providecommand{\noopsort}[1]{}\providecommand{\singleletter}[1]{#1}%

\end{document}